\documentclass{ws-ijmpd}
\usepackage[super,compress]{cite}
\begin{document}

\markboth{J. T. Acu\~{n}a, J. P. Esguerra}
{Dynamics of a planar thin shell at a Taub-FRW junction}

%
\catchline{}{}{}{}{}
%

\title{Dynamics of a planar thin shell at a Taub-FRW junction}

\author{Jan Tristram Acu\~{n}a\footnote{National Institute of Physics, National Science Complex, University of the Philippines Diliman, Quezon City, Philippines}}

\address{National Institute of Physics, University of the Philippines Diliman, Quezon City, 1101, Philippines\\
jacuna@nip.upd.edu.ph}

\author{Jose Perico Esguerra}

\address{National Institute of Physics, University of the Philippines Diliman, Quezon City, 1101, Philippines\\
perryesguerra@gmail.com}

\maketitle

\begin{history}
\received{29 July 2015}
\revised{Day Month Year}
\end{history}

\begin{abstract}
We address the problem of stitching together the vacuum, static, planar-symmetric Taub spacetime and the flat Friedmann-Robertson-Walker cosmology using the Israel thin-shell formalism. The joining of Taub and FRW spacetimes is reminiscent of the Oppenheimer-Snyder collapse used in modeling the formation of a singularity from a collapsing spherical ball of dust. A possible mechanism for the formation of a planar singularity is provided. It is hoped that tackling such example will improve our intuition on planar-symmetric systems in Einstein's general relativity.
\end{abstract}

\keywords{planar symmetric; thin shell; Israel formalism; planar collapse.}

\ccode{PACS numbers: 04.20.Cv}

\section{Introduction}


Similar concepts that appear in different fields in physics are not uncommon. We know from classical electromagnetism (EM) that non-overlapping regions of space must be described by piecewise continuous scalar and vector potentials, but the derivatives of these potentials, i.e. electric and magnetic fields, need not be continuous at the junctions. Discontinuities in the electric and magnetic fields at the junctions give rise to surface charge and current distributions. The same scenario can be seen in classical general relativity (GR) when one tries to patch together non-overlapping regions of spacetime described by different geometries. In this case, the components of the metric tensor are analogous to the 4-potential in classical EM and discontinuities in the derivatives of the metric tensor give rise to surface energy-momentum distributions at the junctions of the spacetimes stitched together. 

A suitable formalism that can be used to patch together spacetimes of different geometries has been formulated by Israel \cite{israel} for the case of non-null hypersurface junctions, and by Barrabes and Israel \cite{barrabes_israel} for the case of null hypersurface junctions. For the case of non-null hypersurface junctions, the conditions for a continuous and smooth transition at the junction between two spacetimes are:
\begin{eqnarray}
\label{1stjunc}\left[h_{ab}\right] &=& 0\\
\label{2ndjunc}\left[K_{ab}\right] &=& 0
\end{eqnarray}
where $h_{ab}$ and $K_{ab}$ are the induced metric and extrinsic curvature on the hypersurface, respectively. In the event that (\ref{2ndjunc}) is not satisfied, a surface energy-momentum distribution exists at the surface junction:
\begin{equation}
S_{ab} = -\frac{\epsilon}{8\pi}\left(\left[K_{ab}\right] - \left[K\right]h_{ab}\right)
\end{equation}
where $K$ is the trace of the extrinsic curvature, and $\epsilon$ takes the value $-1$ for a timelike hypersurface and $+1$ for a spacelike hypersurface. Meanwhile, we also have a similar set of junction conditions for null hypersurface junctions:
\begin{eqnarray}
\label{null1stjunc}\left[\sigma_{AB}\right] &=& 0\\
\label{null2ndjunc}\left[C_{ab}\right] &=& 0
\end{eqnarray}
where $\sigma_{AB}$ is the induced metric of the two-dimensional hypersurface and $C_{ab}$ is the transverse curvature, which is the analog of the extrinsic curvature for null hypersurfaces. If there is a jump in the transverse curvature across the null hypersurface, then there exist surface energy, current, and pressure at the junction.

Israel first used the junction conditions in joining together Schwazrschild and Minkowski spacetimes using a spherical thin shell of pressureless dust. It was found out that: (1) the rest mass of the thin shell is constant throught its history, and (2) the total energy of the shell may be broken down into its kinetic energy and binding energy.\cite{israel} Meanwhile, Barrabes and Israel extended the previous case to a null shell joining an exterior Schwazrschild region and an interior region described by a spherically symmetric spacetime containing a false vacuum. They showed that the expansion of the lightlike shell requires that the energy density of the shell is consumed in the process.\cite{barrabes_israel} 

In the examples mentioned previously, interesting physical interpretations have been drawn in studying those systems. Now, we wish to follow a similar track in studying thin shells in the context of planar symmetric spacetimes. 
Interest in the physical application of planar symmetric systems is rooted from modeling domain walls, which are topological defects that have been formed in spontaneous symmetry breaking processes during cosmological phase transitions. In the thin wall approximation, domain walls may be described by an energy-momentum tensor proportional to a Dirac delta.\cite{vilenkin}   

In this paper, we attempt to consistently patch together the Taub planar symmetric spacetime with a flat Friedmann-Robertson-Walker (FRW) model. This study is motivated by extending the work of Oppenheimer and Snyder in studying the eventual collapse of a spherical ball of dust to a Schwarzschild singularity; we wish to provide a similar mechanism that might explain the formation of planar singularities. 

This paper is organized as follows. Chapter 2 provides a brief account of planar symmetric spacetimes in GR. Chapter 3 contains the main results of this paper; we have demonstrated how to stitch together Taub and FRW spacetimes using subluminal dynamic thin shells. We also looked at certain configurations of dynamic thin shells that will eventually lead to an eternal expansion or collapse of the FRW spacetime. The last chapter contains the conclusions.

\section{The Taub planar symmetric spacetime}
It was shown by Taub \cite{taub1950} that the metric for a vacuum planar-symmetric spacetime with zero cosmological constant can be written as:
\begin{equation}
\label{taub original}ds^2 = \frac{1}{\sqrt{1 + kz}}\left(-dt^2 + dz^2\right) + \left(1 + kz\right)\left(dx^2 + dy^2\right)
\end{equation}
By suitable coordinate transformations, the above may be recast into the following forms\cite{bedranetal, stephanietal}:
\begin{equation}
\label{stephani form}ds^2 = \frac{1}{\sqrt{\zeta}}\left(-dt^2 + d\zeta^2\right) + \zeta\left(dx^2 + dy^2\right),~\zeta > 0
\end{equation}
\begin{equation}
\label{bedran form}ds^2 = -\frac{1}{z^{2/3}}dt^2 + dz^2 + z^{4/3}\left(dx^2 + dy^2\right),~z \neq 0
\end{equation}
One can immediately notice that the Taub spacetime is static since the metric is invariant under translations along the $t$ coordinate. Also, by looking at (\ref{bedran form}), one may ascertain that there is a singularity at $z = 0$. To show that this is a physical singularity, we may evaluate the Kretschmann scalar, and we find that:
\begin{equation}
K \equiv R_{\alpha\beta\gamma\delta}R^{\alpha\beta\gamma\delta} = \frac{64}{27z^4}
\end{equation}
This curvature invariant clearly blows up at $z = 0$, and thus signifies that $z = 0$ is indeed a physical singularity. The hypersurface $z = 0$ has been interpreted by Bedran et. al. as an infinite mass sheet with negative energy density and is repulsive in nature. \cite{bedranetal} One may perform a quick calculation to verify the repulsive nature of this planar singularity. We consider an observer that wishes to be fixed at $z = z_0$. For simplicity, we neglect the motion along the $x$ and $y$ directions. In this case, we may write down the four-velocity of such an observer:
\begin{equation}
u^\alpha \partial_\alpha = |z_0|^{1/3}\partial_t
\end{equation}
The acceleration of this observer in the coordinate basis is:
\begin{equation}
\label{acceleration_taub}a^\mu \partial_\mu = -\frac{1}{3z_0}\partial_\zeta
\end{equation}
Looking at (\ref{acceleration_taub}), one can see that for an observer to stay fixed at $z = z_0 > 0$, the observer must maintain an acceleration pointing towards the planar singularity at $z = 0$, indicating that the planar singularity is repulsive. Similarly, for $z = z_0 < 0$, the observer must maintain an acceleration pointing towards the planar singularity, and we again arrive at the conclusion that the planar singularity is indeed repulsive.

\section{Stitching Taub and FRW spacetimes}
At this point, we want to stitch together Taub and Friedmann-Robertson-Walker (FRW) spacetimes, motivated by the extension of the work of Oppenheimer and Snyder in studying a spherically symmetric collapse of pressureless dust \cite{oppenheimer_snyder} to planar symmetric collapse. In this spherical collapse, the exterior geometry is Schwazrschild while the interior geometry is that of a closed FRW spacetime. A modern treatment by Poisson \cite{poissonbook} showed that a smooth boundary between the two spacetimes implies that the mass of the star $M$ can be related to the energy density $\rho$ of the FRW fluid via:
\begin{equation}
M = \frac{4}{3}\pi \rho R^3
\end{equation}
where $R$ is the radius of the ball of dust.

We have chosen to work with a flat FRW spacetime, whose metric is given by:
\begin{equation}
\label{frw_taubfrw}ds^2 = -dt_-^2 + a^2(t_-)\left(d\zeta^2 + dx^2 + dy^2\right)
\end{equation}
where $a(t_-)$ is known as the scale factor, which accounts for the homogeneous expansion (or collapse) of a Euclidean 3-volume. 

We shall use the Taub metric in describing a static planar symmetric vacuum spacetime:
\begin{equation}
ds^2 = -\frac{1}{z^{2/3}}dt_+^2 + dz^2 + z^{4/3}\left(dx^2 + dy^2\right)
\end{equation}

\subsection{Description of the Planar Junction}
Consider the hypersurface junction $\Sigma$ dividing spacetime into two regions: region I (FRW) and region II (Taub). Region I covers $0 < \zeta < Z_-(\tau)$ and region II covers $z > Z_+(\tau)$. As seen from region II, $\Sigma$ can be parametrized by $t_+ = T_+(\tau)$ and $z = Z_+(\tau)$. The tangent basis vectors on $\Sigma$ as seen from region II are:
\begin{eqnarray}
u^{+\alpha}\partial_\alpha &=& \dot{T}_+\partial_{t_+} + \dot{Z}_+\partial_z\\
e_x^{+\alpha}\partial_\alpha &=& \partial_x\\
e_y^{+\alpha}\partial_\alpha &=& \partial_y
\end{eqnarray}
The overdot stands for differentiation with respect to $\tau$. Hence, the induced metric on $\Sigma$ as seen from region II is:
\begin{equation}
ds_{\Sigma}^2 = -Z_+^{-2/3}\left(\dot{T}_+^2 - Z_+^{2/3}\dot{Z}_+^2\right)d\tau^2 + Z_+^{4/3}(dx^2 + dy^2)
\end{equation}
Meanwhile, as seen from the FRW side, the junction is parametrized by $t_- = T_-(\tau)$ and $\zeta = Z_-(\tau)$. Thus, the tangent basis vectors are:
\begin{eqnarray}
u^{-\alpha}\partial_\alpha &=& \dot{T}_-\partial_{t_-} + \dot{Z}_-\partial_\zeta\\
e_x^{-\alpha}\partial_\alpha &=& \partial_x\\
e_y^{-\alpha}\partial_\alpha &=& \partial_y
\end{eqnarray}
so that the induced metric on $\Sigma$ as seen from region I becomes:
\begin{equation}
ds_{\Sigma}^2 = -\left(\dot{T}_-^2 - a^2(T_-)\dot{Z}_-^2\right)d\tau^2 + a^2(T_-)(dx^2 + dy^2)
\end{equation}

The first junction condition and the normalization condition for 4-velocities imply the following relations:
\begin{eqnarray}
\label{1junc1}\dot{T}_-^2 - a^2(T_-)\dot{Z}_-^2 = Z_+^{-2/3}\left(\dot{T}_+^2 - Z_+^{2/3}\dot{Z}_+^2\right) = 1\\
\label{1junc2}Z_+^{4/3} = a^2(T_-)
\end{eqnarray}
\section{Calculations Regarding the Second Junction Condition}
Let us now calculate the extrinsic curvature on each planar junction. Before we begin, we note that the normal one-form on $\Sigma$ as seen from the FRW side is:
\begin{equation}
\label{normal_frwside}n_\alpha^- dx^\alpha = |a(T_-)|\left(-\dot{Z}_-dt_- + \dot{T}_-d\zeta\right)
\end{equation}
and from the Taub side:
\begin{equation}
\label{normal_taubside}n_\alpha^+ dx^\alpha = \frac{1}{|Z_+|^{1/3}}\left(-\dot{Z}_+dt_+ + \dot{T}_+dz\right)
\end{equation}
Note that we have chosen the normal vector to point from the FRW side to the Taub side, i.e. $n^z > 0$ and $n^\zeta > 0$. The extrinsic curvature can be computed via:
\begin{equation}
K_{ab} \equiv n_{\alpha;\beta}e_a^\alpha e_b^\beta
\end{equation}
Let us begin calculating the extrinsic curvature on the Taub side. We have:
\begin{eqnarray}
K^+_{\tau\tau} &=& n^+_{\alpha;\beta}u^{+\alpha} u^{+\beta}\\
\label{ktt_taubside}&=& -n^+_\alpha a^{+\alpha}
\end{eqnarray}
where $a^{+\mu}$ is the 4-acceleration of an observer comoving with the thin shell in the Taub side. However, 
\begin{eqnarray}
a^{+t} &=& \frac{du^{+t}}{d\tau} + \Gamma^{+t_+}_{\alpha\beta}u^{+\alpha}u^{+\beta}\\
\label{4acct}&=& \ddot{T}_+ - \frac{2}{3Z_+}\dot{T}_+\dot{Z}_+
\end{eqnarray}
and:
\begin{eqnarray}
a^{+z} &=& \frac{du^{+z}}{d\tau} + \Gamma^{+z}_{\alpha\beta}u^{+\alpha}u^{+\beta}\\
\label{4accz}&=& \ddot{Z}_+ - \frac{1}{3Z_+^{5/3}}\dot{T}_+^2
\end{eqnarray}
Thus, using (\ref{normal_taubside}), (\ref{4acct}), and (\ref{4accz}) on (\ref{ktt_taubside}), we have:
\begin{eqnarray}
\nonumber K^+_{\tau\tau} = \frac{\dot{T}_+\text{sgn}(Z_+)}{3Z_+^{4/3}}\left(1 - \dot{Z}_+^2\right)\\
\label{Kplustautau} + \frac{1}{|Z_+|^{1/3}}\left(\dot{Z}_+\ddot{T}_+ - \dot{T}_+\ddot{Z}_+\right)
\end{eqnarray}
Meanwhile, the other non-vanishing components of the extrinsic curvature are:
\begin{eqnarray}
\label{Kspatial_plus}K^+_{xx} = K^+_{yy} = \frac{2}{3}\text{sgn}(Z_+)\dot{T}_+
\end{eqnarray}

We now move on to the FRW side. Similar to the previous calculation of $K^+_{\tau\tau}$ in the Taub side, we first compute the components of the 4-acceleration of the thin shell in the FRW side. We have:
\begin{eqnarray}
a^{-t} &=& \ddot{T}_- + \frac{a(T_-)}{\dot{T}_-}\frac{da(T_-)}{d\tau}\dot{Z}_-^2\\
a^{-z} &=& \ddot{Z}_- + \frac{2}{a(T_-)}\frac{da(T_-)}{d\tau}\dot{Z}_-
\end{eqnarray}
Thus, we have the following component of the extrinsic curvature in the FRW side:
\begin{eqnarray}
\nonumber K^-_{\tau\tau} = |a(T_-)|\left(\dot{Z}_-\ddot{T}_- - \dot{T}_-\ddot{Z}_-\right)\\
\label{Kminustautau} - \text{sgn}(a(T_-))\frac{da(T_-)}{d\tau}\left(\frac{\dot{Z}_-}{\dot{T}_-}\right)\left(1 + \dot{T}_-^2\right)
\end{eqnarray}
Also, we have:
\begin{equation}
\label{Kspatial_minus}K^-_{xx} = K^-_{yy} = a^2(T_-)\text{sgn}(a(T_-))\left(\frac{\dot{Z}_-}{\dot{T}_-}\right)\frac{da(T_-)}{d\tau}
\end{equation}

\subsection{Is a boundary junction possible?}
In the Oppenheimer-Snyder collapse, the surface of the spherical fluid of dust is a boundary layer; a smooth junction exists between the FRW interior and the Schwarzschild exterior. We also ask if this is also the case in stitching Taub and FRW spacetimes. 

For simplicity, we assume that the junction traces a geodesic in either spacetime. Hence, in the Taub side, the 4-velocity is:
\begin{equation}
\label{4veltaub}u^{+\alpha}\partial_\alpha = E\left(Z_+^{4/3}\partial_{t_+} + \sqrt{Z_+^{4/3} - \frac{1}{E^2}}~\partial_z\right)
\end{equation}
where $E$ is a constant of motion due to a $t_-$ translation invariance of the Taub metric. In the FRW spacetime, we have:
\begin{equation}
\label{4velfrw}u^{-\alpha}\partial_\alpha = \sqrt{\frac{p_\zeta^2}{a^2(T_-)} + 1}~\partial_{t_-} + \frac{p_\zeta}{a^2(T_-)}\partial_\zeta
\end{equation}
where $p_\zeta$ is a constant of motion due to a $\zeta$ translation invariance of the FRW metric. Since the trajectory of the junction on either spacetime is a geodesic, comoving observers with the shell experience zero acceleration, and thus $K^+_{\tau\tau} = K^-_{\tau\tau} = 0$. Now, plugging in the components of the 4-velocity of the shell to (\ref{Kspatial_plus}) and (\ref{Kspatial_minus}) and invoking the second junction condition, we arrive at:
\begin{equation}
\frac{2}{3}EZ_+^{2/3} = a(T_-) \frac{p_\zeta~\text{sgn}(a(T_-))}{\sqrt{p_\zeta^2 + a^2(T_-)}}\frac{da(T_-)}{d\tau}
\end{equation}
Using (\ref{1junc2}), we have a differential equation for $a(T_-)$:
\begin{equation}
\label{dea(T)}\frac{2}{3}E = \frac{p_\zeta}{\sqrt{p_\zeta^2 + a^2(T_-)}}\frac{da(T_-)}{d\tau}
\end{equation}
One can show, by using (\ref{dea(T)}) and (\ref{4velfrw}), that the FRW scale factor turns out to be:
\begin{equation}
\label{scale_smooth}a(t_-) = p_\zeta \exp\left[\frac{2E}{3p_\zeta}\left(t_- - T_{-0}\right)\right]
\end{equation}
We have arrived at the conclusion that a smooth junction, where the junction follows geodesic motion in either Taub or FRW spacetime, must have a scale factor given by (\ref{scale_smooth}) at the FRW side. It might be interesting to probe if the system of equations, obtained by invoking the first and second junction conditions, has a unique solution up to certain arbitrary constants, which may be obtained upon setting initial conditions.
\subsection{Friedmann equations in the presence of a boundary layer}
The Friedmann equations may be obtained by plugging in the FRW metric into the Einstein field equations and assuming a cosmological perfect fluid that permeates in FRW spacetime. The Friedmann equations for a flat FRW model 	are:
\begin{eqnarray}
\label{friedmann1}\left(\frac{\dot{a}}{a}\right)^2 &=& \frac{8\pi}{3}\rho\\
\label{friedmann2}\frac{\ddot{a}}{a} &=& -\frac{4\pi}{3}\left(\rho + 3P\right)
\end{eqnarray}
where $\rho$ is the density of the cosmological fluid, $P$ is the pressure of the cosmological fluid, and the overdots in (\ref{friedmann1}) and (\ref{friedmann2}) stand for differentiation with respect to $t_-$. Plugging in (\ref{scale_smooth}) into the Friedmann equations will yield:
\begin{eqnarray}
\rho &=& \frac{3H^2}{8\pi}\\
P &=& -\frac{3H^2}{8\pi}
\end{eqnarray}
where:
\begin{equation}
H \equiv \frac{2E}{3p_\zeta}
\end{equation}
The equation of state $\rho = -P$ for this FRW fluid corresponds to a vacuum-dominated FRW universe. \cite{kolb_turner}

In cosmology, when one models the universe as a flat FRW spacetime containing radiation, matter, and vacuum energy, the evolution of the universe can be marked by different epochs in its expansion: radiation-dominated at early times, then matter-dominated, and finally a vacuum-dominated expansion. \cite{hartle} During a vacuum-dominated expansion, the scale factor is exponentially increasing with FRW coordinate time, i.e. $a(t) \propto e^{Ht}$, and hence ingoing/outgoing photons emitted at some coordinate time $t_a > 0$ will only reach a spatial distance of:
\begin{equation}
d_H \propto \int_{t_a}^\infty \frac{dt'}{a(t')} = \frac{1}{H}e^{-Ht_a}
\end{equation}
Photons cannot reach beyond this spatial distance. This might lead us to suspect that somewhere beyond the cosmological horizon, there exists a boundary layer that separates the FRW spacetime from the Taub spacetime, but cannot be observed from our current location. 
\subsection{Stationary thin shell in FRW}
Let us study the case where the junction is stationary in FRW spacetime, i.e. $\dot{Z}_- = 0$. Even by considering this simple example, it will turn out that we can extract interesting results such as qualitative descriptions regarding the trajectory of the thin shell. Looking at (\ref{Kminustautau}) and (\ref{Kspatial_minus}), we see that the components of the extrinsic curvature of the hypersurface junction as seen from the FRW side vanish, i.e.:
\begin{eqnarray}
K^{-}_{ab} = 0
\end{eqnarray}
Meanwhile, the components of the extrinsic curvature in the Taub side, (\ref{Kplustautau}) and (\ref{Kspatial_plus}), will stay as they are. Notice that unlike the previous case of the thin shell tracing a geodesic trajectory, it appears that the Taub and FRW spacetimes cannot be joined smoothly in this case because if we set $[K_{ab}] = 0$, we will have $\dot{T}_+ = 0$, which will make the 4-velocity of the planar shell to be spacelike in the Taub side. Thus, the planar sheet must be made up of matter whose surface energy-momentum tensor described by:
\begin{eqnarray}
S_{ab} &\equiv& -\frac{1}{8\pi}\left([K_{ab}] - [K]h_{ab}\right)\\
&=& -\frac{1}{8\pi}\left(K^{+}_{ab} - K^{+}h_{ab}\right)
\end{eqnarray}
\begin{eqnarray}
S_{ab} &=& -\left[\begin{array}{ccc}
\frac{\dot{T}_+}{6\pi Z_+^{4/3}}&0&0\\
0&-\frac{\dot{T}_+}{24\pi}\left(\dot{Z}_+^2 + 1\right)&0\\
~&~~~~+\frac{Z_+}{8\pi}\left(\dot{Z}_+ \ddot{T}_+ - \dot{T}_+ \ddot{Z}_+\right)&~\\
0&0&-\frac{\dot{T}_+}{24\pi}\left(\dot{Z}_+^2 + 1\right)\\
~&~&~~~~+\frac{Z_+}{8\pi}\left(\dot{Z}_+ \ddot{T}_+ - \dot{T}_+ \ddot{Z}_+\right)
\end{array}\right]
\end{eqnarray}
If the planar shell is made up of a perfect fluid, then:
\begin{eqnarray}
S_{ab} = \left[\begin{array}{ccc}
\sigma&0&0\\
0&pa^2&0\\
0&0&pa^2
\end{array}\right]
\end{eqnarray}
where $\sigma$ is the surface energy density of the shell and $p$ is the pressure on the shell. Hence, we have:
\begin{eqnarray}
\sigma = -\frac{\dot{T}_+}{6\pi Z_+^{4/3}}\\
\nonumber-p = -\frac{\dot{T}_+}{24\pi Z_+^{4/3}}\left(\dot{Z}_+^2 + 1\right)\\
+ \frac{1}{8\pi Z_+^{1/3}}\left(\dot{Z}_+ \ddot{T}_+ - \dot{T}_+ \ddot{Z}_+\right)
\end{eqnarray}
We may rewrite our expressions for the surface energy density and surface pressure in terms of the scale factor $a(\tau)$. From (\ref{1junc1}) and (\ref{1junc2}), and noting that the thin shell is restricted to be stationary with respect to the interior FRW region, we have:
\begin{eqnarray}
\dot{Z}_+ &=& \frac{3}{2}a^{1/2}(\tau) \dot{a}(\tau)\\
\dot{T}_+ &=& a^{1/2}(\tau) \left(1 + \frac{9}{4}a(\tau)\dot{a}^2(\tau)\right)^{1/2}
\end{eqnarray}
and hence we eventually obtain the surface energy and pressure in terms of the FRW scale factor:
\begin{eqnarray}
\label{sigma_taubfrw}\sigma &=& -\frac{\left(1 + \frac{9}{4}a\dot{a}^2\right)^{1/2}}{6\pi a^{3/2}}\\
\label{p_taubfrw}-p &=& -\frac{\left(1 + \frac{9}{2}a\dot{a}^2 + \frac{9}{2}a^2\ddot{a}\right)}{24\pi a^{3/2}\left(1 + \frac{9}{4}a\dot{a}^2\right)^{1/2}}
\end{eqnarray}
\subsection{Consistency Check}
Now that we have the expressions for the surface energy density and pressure, we wish to check if our expressions for the surface energy density and pressure are correct. The surface energy density and pressure must satisfy the continuity equation:
\begin{equation}
\label{continuity_Sab}S^{ab}_{~~|b} = -\epsilon\left[j^a\right]
\end{equation}
where $j_a = T_{\alpha\beta}e_a^\alpha n^\beta$ is a surface current, and the bar denotes covariant differentiation on the surface, using a connection that is compatible with the induced metric on the junction. The reason why $\sigma$ and $p$ must satisfy (\ref{continuity_Sab}) is because imposing the junction conditions on the junction between the Taub and FRW spacetimes (which are solutions of the EFE) implies that the EFE are satisfied everywhere and, as a result, the energy momentum tensor satisfies the integrability condition $T^{\mu\nu}_{~~;\nu} = 0$ everywhere.

Owing to the fact that the Taub spacetime is a vacuum solution and that the energy-momentum tensor of the FRW spacetime is diagonal, we have $\left[j^a\right] = 0$. Carrying out the calculations, we have:
\begin{eqnarray}
S^{ab}_{~~|b} = \frac{1}{a^2(\tau)}\partial_b\left(a^2(\tau)S^{ab}\right) + \Gamma^a_{bc}S^{bc} &=& 0
\end{eqnarray}
Taking the $a = \tau$ component of the equation, we have:
\begin{eqnarray}
\frac{1}{a^2(\tau)}\frac{d}{d\tau}\left[a^2(\tau)S^{\tau\tau}\right] + \Gamma^\tau_{xx}S^{xx} + \Gamma^\tau_{yy}S^{yy}&=& 0
\end{eqnarray}
Eventually, we obtain:
\begin{eqnarray}
\label{conservation}\frac{d}{d\tau}\left(\sigma a^2\right) + p\frac{d}{d\tau}\left(a^2\right) = 0
\end{eqnarray}
Notice that (\ref{conservation}) has a resemblance with the first law of thermodynamics. To push this resemblance further, we note that the total area of the planar shell is:
\begin{eqnarray}
A &=& \int\int \text{cofac}|h_{\tau\tau}|^{1/2}dx dy\\
&=& a^2(\tau) L^2
\end{eqnarray}
Again, we have introduced a cut-off $L^2$ in place of a divergent integral over all $x$ and $y$. Thus, we identify the ``total energy" of the shell to be $U = \sigma a^2 L^2$ and the ``total area" of the shell to be $A = a^2 L^2$. 

It turns out that the expressions for the energy density and pressure, given by (\ref{sigma_taubfrw}) and (\ref{p_taubfrw}), respectively, satisfy the conservation relation (\ref{conservation}) which is to be expected.
\subsection{Equation of state: $\sigma = -wp$}
Let us consider an equation of state of the form:
\begin{equation}
\sigma = -wp
\end{equation}
for the dynamic thin shell, where $w \neq 0$. Using (\ref{sigma_taubfrw}) and (\ref{p_taubfrw}), we have:
\begin{equation}
\frac{\left(1 + \frac{9}{4}a\dot{a}^2\right)^{1/2}}{6\pi a^{3/2}} = \frac{w}{24\pi a^{3/2}\left(1 + \frac{9}{4}a\dot{a}^2\right)^{1/2}}\left(1 + \frac{9}{2}a\dot{a}^2 +\frac{9}{2}a^2 \ddot{a}\right)
\end{equation}
Simplifying yields:
\begin{equation}
(4 - w) + \frac{9}{2}(2 - w)a\dot{a}^2 - \frac{9}{2}wa^2\ddot{a} = 0
\end{equation}
We may rewrite the above differential equation in terms of $Z_+$, which amounts to the equation of motion of the shell in the spatial direction. From (\ref{1junc2}), again noting that the thin shell junction is stationary relative to the interior FRW region, we now have:
\begin{equation}
(4 - w)\left(1 +\dot{Z}_+^2\right) - 3wZ_+\ddot{Z}_+ = 0
\end{equation}
A straightforward integration of the above expression yields:
\begin{equation}
\label{energy_equation}\dot{Z}_+^2 - \left\vert\frac{Z_+}{Z_{+0}}\right\vert^{\frac{2(4 - w)}{3w}} = -1
\end{equation}
At this point, we may now give qualitative descriptions to the fate of a stationary thin shell junction in the FRW region for any given $w$. One can view the above relation as an effective energy equation with an effective potential:
\begin{equation}
V\left(Z_+\right) = - \left\vert\frac{Z_+}{\tau_0}\right\vert^{\frac{2(4 - w)}{3w}},
\end{equation}
where $\tau_0$ is an integration constant, and energy equal to $-1$. For any $w \neq 0$, there is always a turning point, and hence one can identify the classically allowed and forbidden regions. If the exponent of the effective potential is positive, i.e. $0 < w < 4$, then the thin shell avoids an eventual collapse since the classically allowed region is $[\tau_0, \infty)$, where $\tau_0$ is the location of the turning point. On the other hand, if $w > 4$ or $w < 0$, then the classically allowed region is $(0, \tau_0]$, and hence the thin shell will eventually collapse.
\subsection{Special Case: Domain wall junction}
Supposing that the planar junction is a domain wall, we impose the equation of state $\sigma = -p$.\cite{vilenkin, ipser_skivie, beciu_culetu} A domain wall would correspond to the case $w = 1$. From the previous section, we then see that this domain wall avoids an eventual collapse, and thus the FRW region will expand forever. 
From (\ref{energy_equation}), we may write down the equation of motion of the domain wall:
\begin{equation}
\dot{Z}_+^2 -  \left(\frac{Z_+}{\tau_0}\right)^2 = -1
\end{equation}
Setting $Z_+(0) = Z_0$, we now have a solution for $Z_+(\tau)$:
\begin{equation}
Z_+(\tau) = Z_{+0} \cosh \left(\frac{\tau}{\tau_0} + \eta_0\right)
\end{equation}
where $\eta_0 = \cosh^{-1}\left(\frac{Z_0}{\tau_0}\right)$. We can also rewrite the solution as:
\begin{equation}
Z_+(\tau) = Z_0 \cosh \left(\frac{\tau}{\tau_0}\right) + \sqrt{Z_0^2 - \tau_0^2}\sinh \left(\frac{\tau}{\tau_0}\right)
\end{equation}
Looking at our expression for $Z_+(\tau)$ reinforces our qualitative analysis on the dynamics of the thin shell in the previous section: the thin shell avoids an eventual collapse to a singularity. Meanwhile, the expression for the scale factor $a(\tau)$ will immediately follow:
\begin{equation}
a(\tau) = \left[Z_0 \cosh \left(\frac{\tau}{\tau_0}\right) + \sqrt{Z_0^2 - \tau_0^2}\sinh \left(\frac{\tau}{\tau_0}\right)\right]^{2/3}
\end{equation}
Finally, the surface energy density and pressure of the planar junction are given by:
\begin{equation}
\sigma = -p = -\frac{1}{6\pi\tau_0}\left[1 - \frac{5}{9}\tanh^2\left(\frac{\tau}{\tau_0} + \eta_0\right)\right]^{1/2}
\end{equation}
\subsection{Friedmann equations in the presence of a domain wall junction}
\label{friedmann_domainwall}
Writing:
\begin{equation}
a(\tau) = \tau_0^{2/3} \cosh^{2/3}\left(\frac{\tau}{\tau_0} + \eta_0\right)
\end{equation}
and plugging into the Friedmann equations will yield:
\begin{eqnarray}
\rho &=& \frac{1}{6\pi \tau_0^2}\tanh^2\left(\frac{\tau}{\tau_0} + \eta_0\right)\\
P &=& -\frac{1}{2\pi \tau_0^2}
\end{eqnarray}
Rewriting the energy density of the cosmological fluid so that it contains the pressure of the fluid:
\begin{equation}
\rho = -\frac{1}{3}P\tanh^2\left(\frac{\tau}{\tau_0} + \eta_0\right)
\end{equation}
To aid us in understanding the physical situation, we note that the $z$-coordinate of the planar junction decreases from infinity to $z = Z_0$ in the proper time interval $(-\infty, -\eta_0 \tau_0)$, and then bounces off from $z = \tau_0$ to infinity in the proper time interval $(-\eta_0 \tau_0, \infty)$. In this case, we have an initially contracting FRW spacetime, and then expands after the planar junction reaches the bounce location. Meanwhile, we can see that the density of cosmological fluid in the FRW side decreases from $\frac{1}{6\pi \tau_0^2}$ to zero when the planar junction moves closer to the bounce location, and then increases again asymptotically to $\frac{1}{6\pi \tau_0^2}$ when the planar junction has bounced off $z = \tau_0$.

It can be shown that in the presence of this domain wall junction, the FRW spacetime possesses a cosmological horizon during its expansion phase. Begin by considering the integral
\begin{equation}
d_H = \int_0^\infty \frac{d\tau}{a(\tau)}
\end{equation}
We require $d_H$ to be finite so that this FRW spacetime possesses a cosmological horizon. Now,
\begin{eqnarray}
d_H &=& \tau_0^{-2/3}\int_0^\infty \frac{d\tau}{\cosh^{2/3}\left(\frac{\tau}{\tau_0} + \eta_0\right)}\\
&=& \tau_0^{1/3}\int_{\eta_0}^\infty \frac{du}{\cosh^{2/3}u}\\
&<& \tau_0^{1/3}\int_{-\infty}^\infty \frac{du}{\cosh^{2/3}u}
\end{eqnarray}
Consider the integral:
\begin{equation}
I \equiv \tau_0^{1/3}\int_{-\infty}^\infty \frac{du}{\cosh^{2/3}u}
\end{equation}
Once we have shown $I$ to be finite, it then follows that $d_H$ is bounded above, and hence $I$ tends to a finite value. Letting $x = e^u$ gives:
\begin{equation}
I = \left(4\tau_0\right)^{1/3}\int_0^\infty \frac{x^{-1/3}}{\left(x^2 + 1\right)^{2/3}}dx
\end{equation}
Now, let $x = \tan \theta$. Thus:
\begin{eqnarray}
I &=& \left(4\tau_0\right)^{1/3}\int_0^{\pi/2}\sin^{-1/3}\theta \cos^{-1/3}\theta\\
&=& \left(\frac{\tau_0}{2}\right)^{1/3}B\left(\frac{1}{3},~\frac{1}{3}\right)
\end{eqnarray}
In the last line, we used the integral definition of the Beta function:
\begin{equation}
B(m + 1, n + 1) = 2\int_0^{\pi/2}\cos^{2m + 1}\theta \sin^{2n + 1}\theta d\theta
\end{equation}
Hence, we have the inequality:
\begin{equation}
0 < d_H < \left(\frac{\tau_0}{2}\right)^{1/3}B\left(\frac{1}{3},~\frac{1}{3}\right)
\end{equation}
and thus we have shown that there is a cosmological horizon in this FRW spacetime. Hence, there will come a time when the domain wall will be obscured from view with respect to observers sufficiently far away from the wall.
\section{Conclusions}
We have been able to demonstrate that one can use the Israel thin shell formalism to join together the Taub and FRW spacetimes, and deduce some physical consequences. In stitching the Taub and flat FRW spacetimes, we first obtained expressions for the extrinsic curvature of a generic spacelike hypersurface junction joining these two spacetimes. We have shown that, similar to the Oppenheimer-Snyder collapse, one can have a smooth transition at the junction of these spacetimes. Meanwhile, for a junction that is stationary with respect to the FRW region, the Taub and FRW spacetimes cannot be joined smoothly. As a consequence of the second junction condition, the junction between these spacetimes possesses a surface energy density $\sigma$ and surface tension $-p$, which are dependent on the proper time of an observer comoving with the planar junction. We then considered an equation of state $\sigma = -wp$, and gave qualitative descriptions on the motion of the thin shell for two distinct cases: $0 < w < 4$ (case 1) and $w < 0$, $w > 4$ (case 2). Case 1 does not form a planar singularity, while case 2 eventually leads to a collapse of the thin shell junction to a singularity. We also considered the special case $w = 1$, which corresponds to a domain wall, and obtained an analytic expression for the scale factor and the spatial location of the domain wall. It has been shown that the domain wall cannot form a singularity, which is consistent with our qualitative investigations.

A natural extension to this work is to look for a possible relationship between the equation of state of the surface energy and pressure of the Taub-FRW thin shell junction and the the equation of state of the cosmological fluid. One could also consider general forms for the scale factor $a(\tau)$ and deduce the properties of the thin shell junction. Finally, one could also consider stitching planar symmetric spacetimes other than the vacuum solutions to the field equations.

\section{Acknowledgments}
The authors would like to express their gratitude to Michael Solis and Ian Vega for their comments and suggestions during the preparation of this manuscript. 

\bibliographystyle{ws-ijmpd}
\bibliography{acuna_thinshell_biblio1}
\end{document}